\documentclass[sn-basic]{sn-jnl}

\jyear{2023}%

\theoremstyle{thmstyleone}%
\theoremstyle{thmstyletwo}%

\theoremstyle{thmstylethree}%

\usepackage{natbib}
\usepackage{booktabs}
\usepackage{soul}
\usepackage{amsmath}
\usepackage{amssymb}
\usepackage{makecell}
\usepackage{bbding}
\usepackage{pifont}
\usepackage{bigstrut}
\usepackage{color}
\usepackage{algorithm}
\usepackage{algpseudocode}
\usepackage{multirow}

\begin{document}

\title[Learning Multi-axis Representation in Frequency Domain]{Learning Multi-axis Representation in Frequency Domain for Medical Image Segmentation}


\author[1]{\fnm{Jiacheng} \sur{Ruan}}\email{jackchenruan@sjtu.edu.cn}

\author[1]{\fnm{Jingsheng} \sur{Gao}}\email{gaojingsheng@sjtu.edu.cn}

\author[1]{\fnm{Mingye} \sur{Xie}}\email{xiemingye@sjtu.edu.cn}

\author*[2]{\fnm{Suncheng} \sur{Xiang}}\email{xiangsuncheng17@sjtu.edu.cn}


\affil[1]{\orgdiv{School of Electronic Information and Electrical Engineering}, \orgname{Shanghai Jiao Tong University}, \orgaddress{\city{Shanghai}, \postcode{200240}, \country{China}}}

\affil[2]{\orgdiv{School of Biomedical Engineering}, \orgname{Shanghai Jiao Tong University}, \orgaddress{\city{Shanghai}, \postcode{200240}, \country{China}}}




\abstract{Recently, Visual Transformer (ViT) has been extensively used in medical image segmentation (MIS) due to applying self-attention mechanism in the spatial domain to modeling global knowledge. However, many studies have focused on improving models in the spatial domain while neglecting the importance of frequency domain information. Therefore, we propose \textbf{M}ulti-axis \textbf{E}xternal \textbf{W}eights \textbf{UNet} (\textbf{MEW-UNet}) based on the U-shape architecture by replacing self-attention in ViT with our Multi-axis External Weights block. Specifically, our block performs a Fourier transform on the three axes of the input features and assigns the external weight in the frequency domain, which is generated by our External Weights Generator. Then, an inverse Fourier transform is performed to change the features back to the spatial domain. We evaluate our model on four datasets, including Synapse, ACDC, ISIC17 and ISIC18 datasets, and our approach demonstrates competitive performance, owing to its effective utilization of frequency domain information.}

\keywords{Medical image segmentation, Attention mechanism, Frequency domain information}



\maketitle

\section{Introduction}
\label{sec1}

{M}{ed}ical image segmentation (MIS) has great practical value as it can assist relevant medical staff in locating the lesion area and improve the efficiency of clinical treatment, which has been widely studied, with state-of-the-art methods training convolutional neural networks in a supervised fashion, predicting a label map for a given input image~\citep{dolz2018hyperdense,isensee2021nnu, kamnitsas2017efficient}. For a new segmentation problem, models are typically trained from scratch, requiring substantial design and tuning. Currently, although the CNN-based methods have achieved excellent performance in the field of medical image segmentation, they still cannot fully meet the strict requirements of medical applications for segmentation accuracy. Image segmentation is still a challenge task in medical image analysis area. In recent years, UNet~\citep{unet}, an encoder-decoder model based on U-shape architecture, has been extensively employed for MIS due to its efficient feature extraction capability in handling multi-modal data. To be more specific, the encoder is used to gradually reduce the spatial resolution of the feature map and extract high-level semantic information from the image, while the decoder progressively restores the spatial resolution of the feature map and fuses low-level and high-level features to generate precise segmentation results, which enables UNet to quickly identify fine details in images, even accurately detecting tiny objects. As a result, many studies have been conducted based on the U-shape architecture, such as UNet++~\citep{unet++}, which reduces the semantic gap between the encoder and decoder by incorporating dense connections. Additionally, Att-UNet~\citep{attentionunet} introduces a gating mechanism to enable the model to focus on specific targets.

The above improvements are all based on CNNs, which suffer from poor global information acquisition due to the natural locality of the convolution operation. In contrast, ViT~\citep{vit} incorporates a self-attention mechanism (SA) that enhances the modeling ability of long-range dependencies and effectively captures holistic image semantic information, making it an ideal choice for intensive prediction tasks like image segmentation. As a result, recent advances can be categorized into two types. On the one hand, hybrid structures that combine CNNs and ViTs have gained popularity. For example, UCTransNet~\citep{uctransnet} replaces the skip connection in UNet with the CTrans module, alleviating the issue of incompatible features between the encoder and decoder. MT-UNet~\citep{mtunet} employs CNNs at a shallow level and Local-Global SA along with the external attention mechanism at a deep level, to obtain richer representation information. On the other hand, some studies have leveraged pure ViTs, such as Swin-UNet~\citep{swinunet}, which replaces the convolution operation in U-Net with the Swin Transformer Block and achieves superior results. Although the aforementioned models focusing on spatial domain have achieved notable success in medical image area or other vision area~\citep{ruan2024vm,xiang2022rethinking,xiang2023learning,xiang2020unsupervised1}, the spatial domain modeling generally suffers
from the low efficiency, e.g. it is very difficult to linearize the complexity of the vision Transformer from the perspective of spatial domain, which make them less favored by many image downstream tasks.

\begin{figure}[!t]
\centering
\includegraphics[width=1.00\columnwidth]{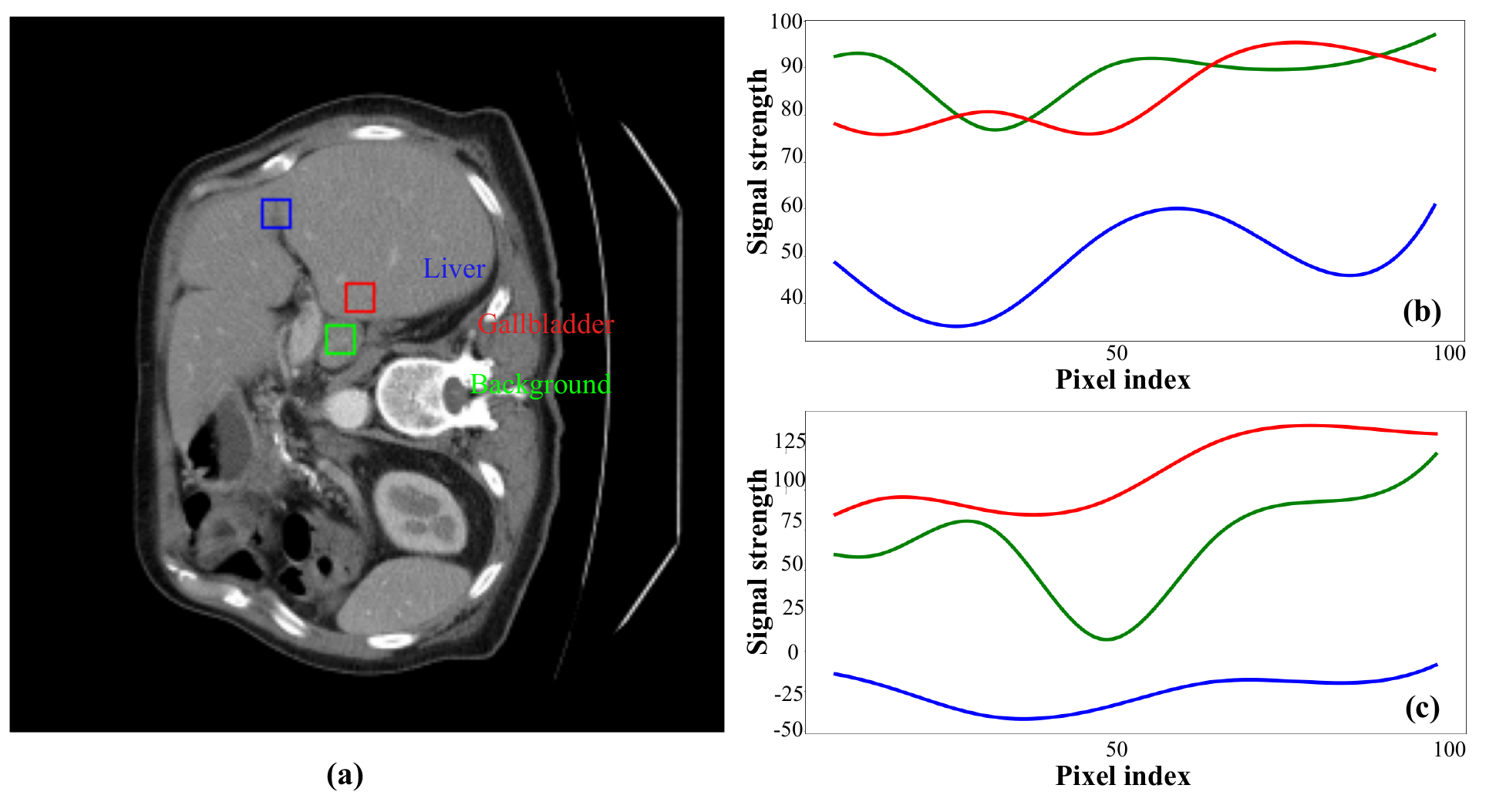}
\caption{Frequency-aware clues for medical image segmentation. (a) Discrete Fourier Transform (DFT) is used in three $10\times10$ patches. \textcolor{blue}{Blue} represents segmentation region 1 (Liver), \textcolor{red}{red} represents segmentation region 2 (Gallbladder), and \textcolor{green}{green} represents background. (b) the frequency signal strength curve of selected patches when DFT is performed only on a single axis. (c) performs DFT on three axes comprehensively.}
\label{fig1}
\end{figure}

In essence, the existing medical image segmentation methods mainly include spatial technique and frequency technique, among which spatial technique has some disadvantages. However, frequency domain technology has improved these problems.
In general computer vision, feature extraction in the frequency domain has been shown to be a powerful approach~\citep{frequencyCOD,gfnet} to achieve more efficient and effective encoding and decoding processes. For instance, GFNet~\citep{gfnet} leverages 2D discrete Fourier transform (2D DFT) to convert features from the spatial domain to the frequency domain, and filters in the frequency domain are used for learning representation. For MIS, many pathological areas are difficult to extract in the spatial domain, while they can be more easily extracted in the frequency domain~\citep{bibmMedicalFrequencyDomainLearning}. However, previous methods extract the frequency domain information only in a single axis, resulting in some signals that are still difficult to distinguish, which can be explained in Figure~\ref{fig1}. In Figure~\ref{fig1} (a), it is challenging to visually distinguish the three regions from the spatial domain perspective. In contrast, Figure~\ref{fig1} (b) illustrates the signal strength of the three regions in the frequency domain, revealing relatively apparent differences in signal strength among them. However, there are still some signal strength intersections between the \textcolor{green}{\textbf{green}} and \textcolor{red}{\textbf{red}} curves in the case of employing DFT only for a single axis (the Height-Width axis). By considering the frequency domain signal strength of the three axes together, as shown in Figure~\ref{fig1} (c), it becomes evident that there is no signal strength intersection among the three curves. Motivated by this observation, we propose to extract and fuse features using a multi-axis approach.

Based on the aforementioned findings, we propose the Multi-axis External Weights mechanism (MEW) that can simultaneously capture more comprehensive global and local information. To be specific, the feature map is partitioned into four branches along the channel dimension. For the first three branches, the features are transformed into the frequency domain using 2D DFT along three different axes (the Height-Width, Channel-Width and Channel-Height axes). Subsequently, the corresponding learnable weights are utilized to multiply the frequency domain maps to obtain the frequency domain information and global knowledge. In addition, depthwise separable (DW) convolution operation~\citep{mobilenetv2} is conducted for the remaining branch to obtain local information. Thereafter, MEW is integrated into the ViT by replacing the SA module, resulting in the Multi-axis External Weights Block (MEWB). Finally, based on MEWB and U-shape architecture, a powerful network for medical image segmentation called MEW-UNet is proposed. 

To this end, the major contributions of our work can be summarized as follows:
\begin{itemize}
\item[$\bullet$] The Multi-Axis External Weights block is proposed for the first time to obtain global and local information simultaneously, and frequency domain feature signals are introduced to comprehend the context more effectively.

\item[$\bullet$] Based on U-shape and ViT's structure, we replace the self-attention block in ViT with our proposed block, resulting in a powerful MIS model, dubbed MEW-UNet.

\item[$\bullet$] We conduct comprehensive experiments on four datasets and achieve state-of-the-art results, demonstrating the efficacy of our methods.
\end{itemize}
The remainder of this paper is structured as follows. In Section~\ref{sec2}, we give the related works based on multi-axis representation learning in medical image segmentation, and then briefly introduce our method.
In Section~\ref{sec3}, The details of our multi-axis external weights UNet based on the U-shape architecture are presented. Extensive evaluations compared with state-of-the-art methods and comprehensive
analyses of the proposed approach are reported in Section~\ref{sec4}. Finally, conclusion of this paper and discussion of future works are presented in Section~\ref{sec5}.

\section{Related Work}
\label{sec2}

\subsection{Medical Image Segmentation}
In essence, medical image processing requires extremely high accuracy for disease diagnosis~\citep{chen2023colo}. Segmentation in medical imaging refers to pixel-level or voxel-level segmentation. Generally, the boundary between multiple cells and organs is difficult to be distinguished on the image. 

First, a lot of  endeavours  have  been  concentrating  on the  segmentation  process. These methods  conquer  different  restrictions  on conventional medical  division  techniques.  Yet,  there  isn't so  much  as one single strategy to be considered a better technique for different kinds of images; those techniques are only suitable for particular images and other applications~\citep{xiang2023less,gao2024lamm}.

Second, medical images are acquired from various medical equipment and the standards for them and annotations or performance of CT/MR3 machines are not uniform. Hence deep-learning-related trained models are only suitable for specific scenarios~\citep{xiang2023rethinking}. Meanwhile, the deep network with weak generalization may easily capture wrong features from the analyzed medical images. Furthermore, significant inequality always exists between the size of negative and positive samples, which may have a greater impact on the segmentation. However, U-Net could afford an approach achieving better performance in reducing the dilemma of overfitting problem.

\subsection{ViT Based Techniques}
Transformer models~\citep{dosovitskiy2020image} entirely rely on the self-attention mechanism to build long-distance dependencies, which have achieved great success in almost all natural language processing tasks~\citep{gao2023lamm}. Vision transformer (ViT)~\citep{xiang2023deep} is one of the earlier attempts to introduce transformer models into vision tasks, which applies a pure transformer architecture on non-overlapping image patches for image classification and has achieved state-of-the-art accuracy. Since ViT models excel at capturing spatial information, they have also been extended to more challenging tasks, including action detection~\citep{li2023av}, image retrieval~\citep{xiang2020progressive} and segmentation~\citep{you2023autokary2022}. For example, UCTransNet~\citep{wang2022uctransnet} replaced the skip-connection with the channel transformer (CTrans) module. \citep{karimi2021convolution} applied self-attention between neighboring image
patches by modifying the MHSA mechanism of vision transformers. As for the model design, \cite{duan2023dynamic} proposed a hybrid transformer
named DUCT consisting of
dynamic local enhancement, unary co-occurrence
excitation, and a standard multi-head self-attention module to learn the local, mid-level
and global information, respectively. And a deep semi-supervised HAR approach named MixHAR~\citep{duan2024wearable} is also introduced to employ a linear interpolation mechanism to blend labelled and unlabelled activities while addressing both inter- and intra-activity variability.
Despite the excellent performance in multiple image segmentation tasks. Vision Transformers suffer from the problem of computational overload, which has never been solved.

To address this problem, in this work, we make a first attempt to propose Multi-Axis External Weights block to obtain the frequency domain information on medical image. To the best of our knowledge, this is the first attempt to leverage the Multi-axis External Weights mechanism (MEW) that can simultaneously capture more comprehensive global and local information. We hope this work could shed some new light on potential tasks for medical image community to move forward to move forward.

\section{Our method}
\label{sec3}

\subsection{Preliminary}
The problem of medical image segmentation can often be posed as one of optimization of an appropriately defined objective function. The objective function is usually complex, multimodal, discontinuous, and cannot be described in a closed mathematical form that can be analytically solved. In this work, we make a new attempt and propose a novel multi-axis external weights block for global and local information. On the basis of it, the frequency domain feature signals are introduced to help the model fully understand the context. On the whole, the goal of this paper is to leverage frequency domain information to learn discriminative embeddings for downstream medical image segmentation task. An overview of the proposed architecture is illustrated in Figure~\ref{fig2}.

\subsection{Multi-axis External Weights Block}

\begin{figure}[!t]
	\centerline{\includegraphics[width=1.00\textwidth]{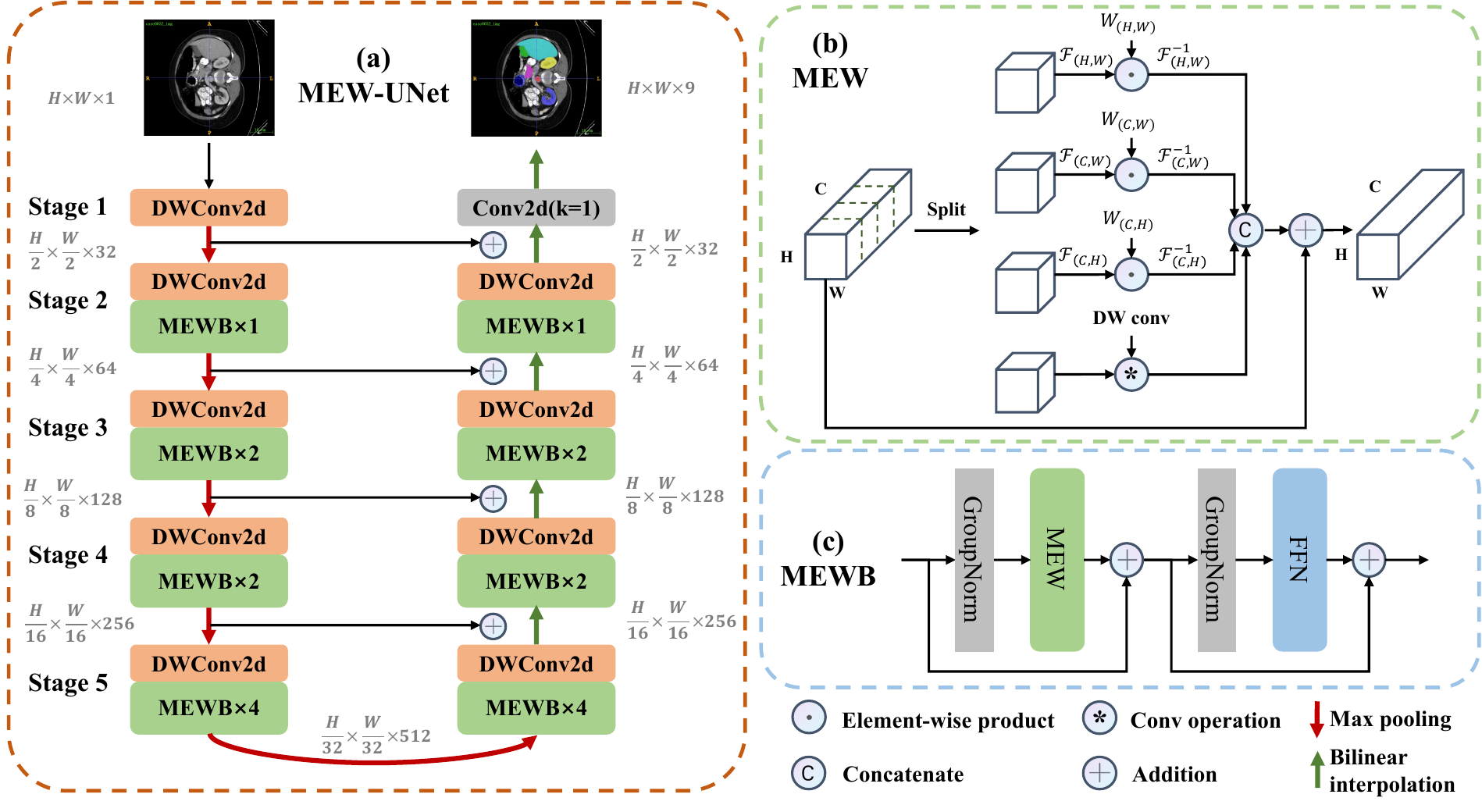}}
	\caption{(a) The overall architecture of MEW-UNet. (b) Multi-axis External Weights mechanism. $\mathcal{F}_{(H,W)}$ and $\mathcal{F}^{-1}_{(H,W)}$ refer to conducting 2D DFT and 2D inverse DFT along the Height-Width axis of the feature map. $\mathcal{F}_{(C,W)}$, $\mathcal{F}^{-1}_{(C,W)}$, $\mathcal{F}_{(C,H)}$, and $\mathcal{F}^{-1}_{(C,H)}$ could be illustrated as the same. (c) Multi-axis External Weights block. FFN presents the feed-forward layer.}
\label{fig2}
\end{figure}

In MIS, recent methods have been primarily focused on obtaining information in the spatial domain, while neglecting the importance of the frequency domain. In the spatial domain, the boundary between the segmented object and the background is often indistinct, whereas in the frequency domain, the objects are at distinct frequencies, which can be readily differentiated~\citep{bibmMedicalFrequencyDomainLearning}. Although the idea of utilizing the frequency domain has been introduced in prior works such as~\citep{bibmMedicalFrequencyDomainLearning} and~\citep{gfnet}, extracting frequency domain features on a single axis prevents the model from further distinguishing segmentation boundaries clearly, leading to degraded performance. Therefore, we propose MEW, which is based on 2D DFT for different axes, to obtain more comprehensive information in the frequency domain, as shown in Figure~\ref{fig2} (b).

An input feature map $X\in\mathbb{R}^{C\times H\times W}$ with $C$, $H$, and $W$ representing the channel, height, and width of the map is considered. $X$ is initially divided into four equal parts along the channel dimension, which are subsequently sent into four different branches. The MEW mechanism is expressed by the Eq.~\ref{formula1} to Eq.~\ref{formula4}.

\begin{equation}
	x_{1}, x_{2}, x_{3}, x_{4}=Split(X)
\label{formula1}
\end{equation}

\begin{equation}
	x_{i(I,J)}=W_{(I,J)} \odot \mathcal{F}_{(I,J)}[x_i] 
\label{formula2}
\end{equation}

\begin{equation}
x_{i}^\prime = \mathcal{F}^{-1}_{(I,J)}[x_{i(I,J)}], x_{4}^\prime = DW(x_4)
\label{formula3}
\end{equation}

\begin{equation}
Y = Concat(x_{1}^\prime,x_{2}^\prime,x_{3}^\prime,x_{4}^\prime) + X
\label{formula4}
\end{equation}
where $i = 1,2,3$ corresponds to the first three branches. $W_{(I,J)}$ and $\mathcal{F}_{(I,J)}$ denote the learnable external weight and the 2D DFT for the corresponding axes. DW represents the depthwise separable convolution operation. When $i = 1$, $I$ and $J$ represent Height-Width axes. When $i = 2$, $I$ and $J$ present Channel-Width axes. When i = 3, $I$ and $J$ represent Channel-Height axes. $\odot$ is the element-wise product. $\mathcal{F}^{-1}_{(I,J)}$ refers to the 2D inverse DFT. $Split$ and $Concat$ represent the split and concatenation operation along the channel dimension respectively.

In the first branch, the feature map is transformed to the frequency domain by performing 2D DFT on the Height-Width axes. Subsequently, the element-wise product operation is conducted between the feature map and the corresponding learnable external weight. Finally, the map is converted back to the spatial domain by applying the 2D inverse DFT. In the same way, the second and third branches perform the above operations on the Channel-Width and Channel-Height axes, respectively. This multi-axis operation enables more comprehensive global information to be learned. It should be noted that the external weight is generated by our External Weights Generator, as explained in the next section. In addition, local information is also critical for the MIS task. Therefore, for the fourth branch, the DW convolution is utilized to obtain local information. Subsequently, the feature maps of the four branches are concatenated along the channel dimension to restore the same size as the input. Finally, the residual connection of the input is employed to obtain the output.

As shown in Figure \ref{fig2} (c), we replace SA in ViT with our MEW and obtain our MEWB, which can be expressed as Eq.~\ref{formula5} and Eq.~\ref{formula6} respectively.
\begin{equation}
	X' = MEW(GroupNorm(X)) + X
\label{formula5}
\end{equation}
\begin{equation}
	Y = FFN(GroupNorm(X')) + X'
\label{formula6}
\end{equation}
where MEW denotes the multi-axis external weights block and $GroupNorm$ represents the group normalization~\citep{gn}. Besides, $FFN$ indicates the feed-forward layer.
It should be noted that we adopt GroupNorm instead of LayerNorm in the vanilla ViT. This is due to the fact that we directly utilize the input $X \in \mathbb{R}^{C \times H \times W}$ instead of performing patch embedding operation. Furthermore, medical image datasets often have small sample sizes, and the batch size used in experiments is also typically small. In our network, we apply GroupNorm with four groups to mitigate the influence of batch size in BatchNorm~\citep{bn}.

\subsection{External Weights Generator}

MIS is classified as a layout-specific task, where the variation between samples is minor but within samples is significant in a particular medical dataset~\citep{tvconv}. Merely relying on the randomly initialized learnable weights cannot effectively build the semantic relationship between different regions. Hence, this paper introduces an External Weights Generator to transform the initialized weights. Our generator can be expressed as Eq.~\ref{formula7}.

\begin{equation}
	W_{(I,J)} = IRB(BI(W_{(I,J)}^{init}))
\label{formula7}
\end{equation}
where $W_{(I,J)}^{init}$ presents the initial learnable tensor. $BI$ means a bilinear interpolation operation. $IRB$ refers to Inverted residual blocks~\citep{mobilenetv2}.

Specifically, the External Weights Generator comprises several DW convolutions, which are utilized to generate the weights corresponding to the Height-Width, Channel-Width, and Channel-Height axes. For instance, to generate the weight for the Height-Width axis, a learnable tensor is randomly initialized, which is then subjected to a bilinear interpolation operation. Subsequently, the learnable tensor undergoes the Inverted residual blocks to obtain the final weight, which is used in the element-wise product operation (referring to Eq.~\ref{formula2} for better understanding). Similarly, the weights for the Channel-Width and Channel-Height axes are obtained through the same process.

\section{Experiments}
\label{sec4}

\subsection{Datasets}
\label{sec4.1}

We evaluate our method on three benchmark datasets, including ISIC17~\citep{isic17}, ISIC18~\citep{isic18}, Synapse~\citep{synapse} and ACDC~\citep{acdc}. The evaluation protocols are also elaborated before showing our experiments results.

\textbf{ISIC17}~\citep{isic17} and \textbf{ISIC18}~\citep{isic18}, two public datasets of skin lesion segmentation, containing 2150 and 2694 dermoscopy images with ground truth, respectively. We randomly divide datasets in a ratio of 7:3 for training and testing. Five evaluation metrics are reported, including Mean Intersection over Union (mIoU), Dice Similarity Coefficient (DSC), Accuracy (Acc), Sensitivity (Sen), and Specificity (Spe).

\textbf{Synapse}~\citep{synapse}, a public multi-organ segmentation dataset, consists of 30 abdominal CT cases. Following~\cite{transunet}, we use 18 cases for training and 12 cases for testing. Dice Similarity Coefficient (DSC) and 95\% Hausdorff Distance (HD95) are used to evaluate our method on this dataset.

\textbf{ACDC}~\citep{acdc}, a public cardiac MRI dataset, is composed of 100 MRI scans. Following \citep{mtunet}, we use the same split of 70 training cases, 10 validation cases and 20 testing cases. Dice Similarity Coefficient (DSC) and 95\% Hausdorff Distance (HD95) are reported as metrics.

\subsection{Implementation details}

Following previous works~\citep{malunet, mtunet}, for the ISIC17 and ISIC18 datasets, all images are resized to 256 × 256. For the Synapse and ACDC dataset, all images are resized to 224 × 224. To avoid overfitting, data augmentation is performed, including random flip and random rotation. The loss function is BceDice loss. We set batch size equal to 8, and utilize CosineAnnealingLR~\citep{cosineannealingLR} as the scheduler. Following the parameter setting in \cite{adamw}, the initial learning rate (lr), maximum epochs (ep), and optimizer (opt) for the three datasets are itemized below:
\begin{itemize}
	\setlength{\itemsep}{0pt}
	\setlength{\parsep}{0pt}
	\setlength{\parskip}{0pt}
	\item ISIC17 and ISIC18: lr=1e-3; ep=300; opt=AdamW; 
	\item Synapse: lr=3e-3; ep=600; opt=SGD.
\end{itemize}

Besides, the same settings as~\citep{mtunet} are conducted for the ACDC dataset, except that we set batch size equal to 4. All experiments are conducted on a single NVIDIA RTX A6000 GPU.

\begin{table}[!t]
\centering
\caption{Comparative experimental results on the ISIC17 and ISIC18 dataset.}
\small
\setlength{\tabcolsep}{0.77mm}{
\begin{tabular}{c|c|cc}
\hline
\textbf{Dataset} &\textbf{Model}          & \textbf{mIoU(\%)$\uparrow$}  & \textbf{DSC(\%)$\uparrow$}   \\ \hline
\multirow{6}{*}{ISIC17} &UNet~\citep{unet}      & 76.98    & 86.99    \\
&UTNetV2~\citep{utnetv2}       & 77.35     & 87.23       \\
&TransFuse~\citep{transfuse}     & 79.21     & 88.40  \\
&MALUNet~\citep{malunet} & 78.78 & 88.13 \\
&EGE-UNet~\citep{egeunet} & 79.81 & 88.77 \\
&\textbf{MEW-UNet (Ours)} & \textbf{81.38} & \textbf{89.73}   \\ \hline
\multirow{9}{*}{ISIC18}&UNet \cite{unet}                    & 77.86          & 87.55            \\
&UNet++~\citep{unet++}                 & 78.31          & 87.83          \\
&Att-UNet~\citep{attentionunet}               & 78.43          & 87.91        \\
&UTNetV2~\citep{utnetv2}                & 78.97          & 88.25         \\
&SANet~\citep{sanet}                  & 79.52          & 88.59        \\
&TransFuse~\citep{transfuse}     & 80.63          & 89.27    \\
&MALUNet~\citep{malunet} & 80.25 & 89.04 \\
&EGE-UNet~\citep{egeunet} & 80.94 & 89.46 \\
&\textbf{MEW-UNet (Ours)} & \textbf{81.90} & \textbf{90.05} \\ \hline
\end{tabular}}
\label{tab1}
\end{table}

\subsection{Comparison with State-of-the-Arts}

We compare our model with State-of-the-Arts in recent years, such as MT-UNet~\citep{mtunet}, UCTransNet~\citep{uctransnet}, TransFuse~\citep{transfuse}, SANet~\citep{sanet}, and so on. The experimental results are shown in Table~\ref{tab1}, Table~\ref{tab2} and Table~\ref{tab2_2}. For the ISIC17 and ISIC18 datasets, our MEW-UNet outperforms all other state-of-the-arts on the mIoU and DSC metrics. For the Synapse dataset, our MEW-UNet is 0.33\% and 0.69\% higher than MT-UNet and UCTransNet in terms of DSC. Besides, it is worth noting that our model surpasses MT-UNet and UCTransNet \textbf{10.15mm} and \textbf{10.31mm} in the aspect of HD95. For the ACDC dataset, our MEW-UNet also surpasses MT-UNet in terms of DSC and HD95 metrics. 

\begin{table}[!t]
\centering
\caption{Comparative experimental results on the Synapse dataset.}
\small
\setlength{\tabcolsep}{2.23mm}{
\begin{tabular}{ccc}
\hline
\multicolumn{3}{c}{\textbf{Synapse dataset}}    \\
\hline
\multicolumn{1}{c|}{\textbf{Model}}           & \textbf{\textbf{DSC(\%)$\uparrow$}}   & \textbf{\textbf{HD95(mm)$\downarrow$}}  \\
\hline
\multicolumn{1}{c|}{V-Net~\citep{vnet}}                    & 68.81          & -                     \\
\multicolumn{1}{c|}{UNet~\citep{unet}}                     & 76.85          & 39.70               \\
\multicolumn{1}{c|}{Att-UNet~\citep{attentionunet}  }                 & 77.77          & 36.02           \\
\multicolumn{1}{c|}{TransUnet~\citep{transunet}}                & 77.48          & 31.69                  \\
\multicolumn{1}{c|}{UCTransNet~\citep{uctransnet}  }               & 78.23          & 26.75              \\
\multicolumn{1}{c|}{MT-UNet~\citep{mtunet} }                  & 78.59          & 26.59               \\
\multicolumn{1}{c|}{\textbf{MEW-UNet (Ours)}} & \textbf{78.92} & \textbf{16.44}     \\
\hline
\end{tabular}}
\label{tab2}
\end{table}

\begin{table}[!t]
\centering
\caption{Comparative experimental results on the ACDC dataset.}
\small
\setlength{\tabcolsep}{3.07mm}{
\begin{tabular}{ccc}
\hline
\multicolumn{3}{c}{\textbf{ACDC dataset}}    \\
\hline
\multicolumn{1}{c|}{\textbf{Model}}           & \textbf{\textbf{DSC(\%)$\uparrow$}}   & \textbf{\textbf{HD95(mm)$\downarrow$}}    \\
\hline
\multicolumn{1}{c|}{R50 UNet~\citep{transunet} }                 & 87.60          & -                 \\
\multicolumn{1}{c|}{R50 AttnUNet~\citep{transunet} }             & 86.90          & -                 \\
\multicolumn{1}{c|}{R50 ViT~\citep{transunet} }                  & 86.19          & -                 \\
\multicolumn{1}{c|}{TransUNet~\citep{transunet} }                & 89.71          & -                 \\
\multicolumn{1}{c|}{Swin-Unet~\citep{swinunet}  }                & 88.07          & -                 \\
\multicolumn{1}{c|}{MT-UNet~\citep{mtunet}}                  & 90.43          & 2.23              \\
\multicolumn{1}{c|}{\textbf{MEW-UNet (Ours)}} & \textbf{91.00} & \textbf{1.19}     \\ \hline
\end{tabular}}
\label{tab2_2}
\end{table}

\subsection{Ablation Studies}

\begin{table}[!t]
\centering
\caption{Ablation Studies for our MEWB on the ISIC18 dataset. $Random$ represents only randomly initialized learnable weights without $IRB$.}
\small
\renewcommand\arraystretch{1.25}
\setlength{\tabcolsep}{1.27mm}{
\begin{tabular}{ccccc|cc}
\hline
DW   & \textbf{$W_{(H, W)}$}   & \textbf{$W_{(C, W)}$}  & \textbf{$W_{(C, H)}$} &\textbf{$Random$}   & \textbf{mIoU(\%)$\uparrow$} & \textbf{DSC(\%)$\uparrow$} \\ \hline
\textbf{\checkmark} &  &   &    &     & 80.14       & 88.98    \\
\textbf{\checkmark} & \textbf{\checkmark} & & &  & 80.82   & 89.39  \\
\textbf{\checkmark} & \textbf{\checkmark} & \textbf{\checkmark} &          &                                           & 81.27                                    & 89.67                                   \\
& \textbf{\checkmark} & \textbf{\checkmark} & \textbf{\checkmark} & & 81.29                                    & 89.68                                   \\
\textbf{\checkmark} & \textbf{\checkmark} & \textbf{\checkmark} & \textbf{\checkmark} &  & 81.90                                    & 90.05                                   \\
\textbf{\checkmark} & \textbf{\checkmark} & \textbf{\checkmark} & \textbf{\checkmark} & \textbf{\checkmark} & 80.59                                    & 89.25                                   \\ \hline
\end{tabular}}
\label{tab:ablonisic}
\end{table}

\begin{table}[!t]
\centering
\caption{Ablation Studies for our MEWB on the ACDC dataset. $Random$ represents only randomly initialized learnable weights without $IRB$.}
\small
\renewcommand\arraystretch{1.25}
\setlength{\tabcolsep}{1.27mm}{
\begin{tabular}{ccccc|cc}
\hline
DW   & \textbf{$W_{(H, W)}$}   & \textbf{$W_{(C, W)}$}  & \textbf{$W_{(C, H)}$} &\textbf{$Random$}   & \textbf{DSC(\%)$\uparrow$} & \textbf{HD95(mm)$\downarrow$} \\ \hline
\textbf{\checkmark} &  &   &    &     &   90.05   & 1.65    \\
\textbf{\checkmark} & \textbf{\checkmark} & & &  &  90.22  &  1.30 \\
\textbf{\checkmark} & \textbf{\checkmark} & \textbf{\checkmark} &          &                                           &  90.44                                  &              1.25                     \\
& \textbf{\checkmark} & \textbf{\checkmark} & \textbf{\checkmark} & &              90.54                      &               1.22                    \\
\textbf{\checkmark} & \textbf{\checkmark} & \textbf{\checkmark} & \textbf{\checkmark} &  &       91.00                              &       1.19                             \\
\textbf{\checkmark} & \textbf{\checkmark} & \textbf{\checkmark} & \textbf{\checkmark} & \textbf{\checkmark} &           90.23                          &                  1.29                 \\ \hline
\end{tabular}}
\label{tab:ablonacdc}
\end{table}

The fundamental concept behind our MEW-UNet architecture lies in its multi-axis operations. To evaluate the effectiveness of this approach, we gradually increase the number of multi-axis operations and analyze the performance in an ablation study, as presented in Table~\ref{tab:ablonisic}. It can be observed that the adoption of DW convolution alone results in a decrease of almost 2\% in the mIoU metric. Similarly, when solely utilizing single-axis frequency-domain transformation, the mIoU metric also decreases by 1\%. The results indicate that our multi-axis operations can effectively capture comprehensive frequency domain information and global knowledge. Additionally, even when multi-axis operations are utilized, there is still a slight degradation in performance when leaving out the DW convolution in the fourth branch, which further emphasizes the importance of local information. On the other hand, the same phenomenon can also be observed when our Multi-axis External Weights block is tested on the ACDC datasets. As illustrated in Table~\ref{tab:ablonacdc}, there also exists obvious performance degradation of 0.78\% in terms of Dice Similarity Coefficient when only applied with the single-axis frequency-domain transformation. This confirms that frequency domain information is a crucial ingredient for increasing performance
to an excellent level.
Finally, we demonstrate the efficacy of our External Weights Generator by only randomly initializing learnable weights without IRB, and the degradation in performance can be clearly seen, which underscores the importance of using IRB for transforming learnable weights.

\subsection{Visualization}
\begin{figure}[!t]
    \centering
    \includegraphics [width=1.00\textwidth] {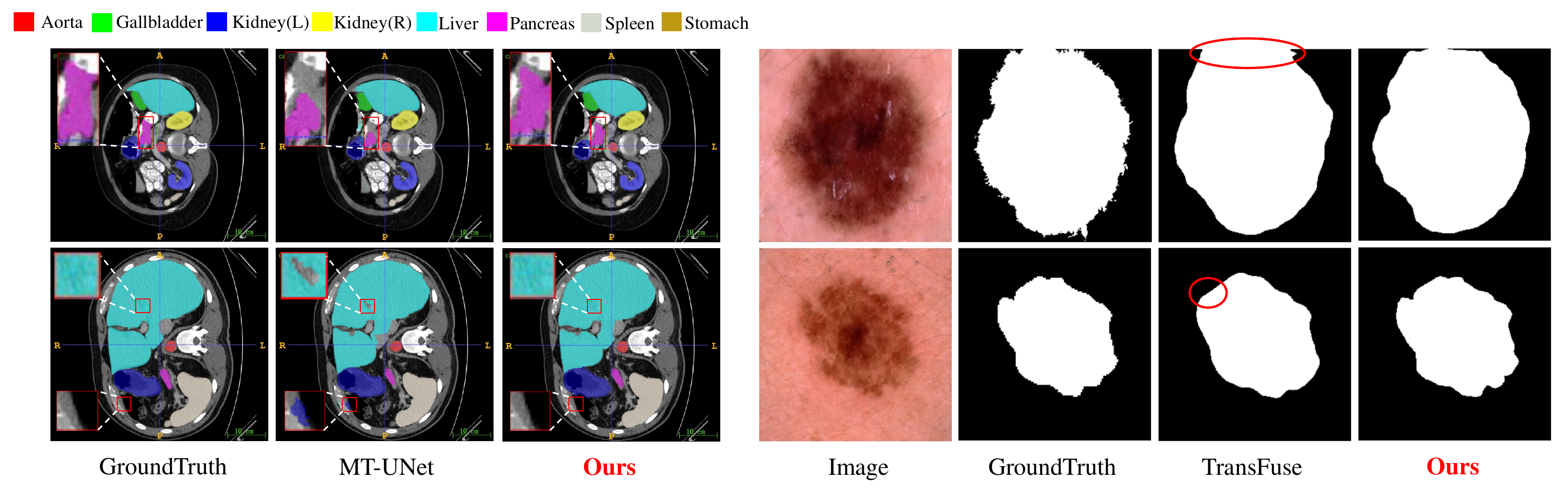}
	\caption{The visualization comparison on the Synapse and ISIC2018 datasets.}
    \label{fig3}
\end{figure}

\begin{figure}[!t]
    \centering
    \includegraphics [width=1.00\textwidth] {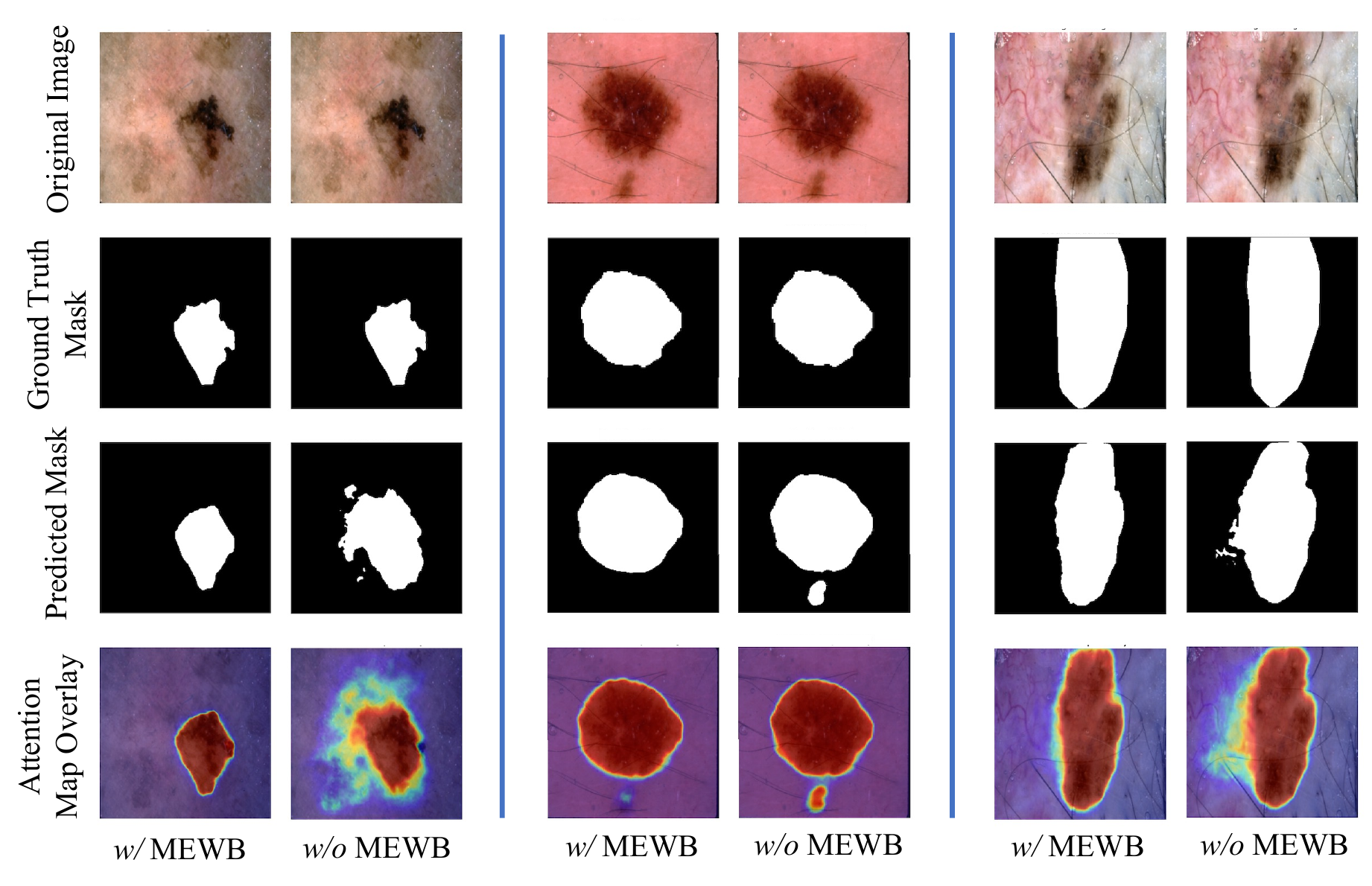}
	\caption{The visualization comparison of the proposed Multi-axis External Weights Block on ISIC2018 dataset.}
    \label{fig:attmaps}
\end{figure}

To go even further, we also give some qualitative results of our proposed Multi-axis External Weights UNet named MEW-UNet.
Figure~\ref{fig3} provides some visualized segmentation results by ITK-SNAP~\citep{ITK-SNAP}. For instance, our MEW-UNet has on hole and missing boundary compared with MT-UNet on the Synapse dataset. Besides, compared to TransFuse~\citep{transfuse}, our model has more clearly boundary on the ISIC2018 dataset on this challenging segmentation task. On the other hand, we also visualize the attention feature maps of the proposed Multi-axis External Weights Block (MEWB) on the ISIC2018 dataset, according to the Figure~\ref{fig:attmaps}, we can easily observe that the attention maps with MEWB demonstrate a more precise focus on the lesion regions compared to the ones without MEWB. The attention overlay highlights how the model with MEWB captures the intricate details and boundaries of the lesions more effectively, leading to improved segmentation performance.

\section{Conclusion}
 \label{sec5}

In this paper, we propose a Multi-axis External Weights block to acquire frequency domain information more comprehensively. Our External Weight Generator is applied to obtain learnable weights for different axes. Furthermore, we replace the self-attention block with our block to construct MEW-UNet. Experimental results demonstrate that our model achieves state-of-the-art performance. We believe that our work will provide new insights for subsequent model developments in the frequency domain. In the future, we plan to extend our multi-axis design to other medical tasks.

\bmhead{Acknowledgments}

This work was partially supported by the National Natural Science Foundation of China under Grant No.62301315, Startup Fund for Young Faculty at SJTU (SFYF at SJTU) under Grant No.23X010501967 and Shanghai Municipal Health Commission Health Industry Clinical Research Special Project under Grant No.202340010.
The authors would like to thank the anonymous reviewers for their valuable suggestions and constructive criticisms.

\section*{Declarations}

\begin{itemize}
\item \textbf{Funding} \\  This work was partially supported by the National Natural Science Foundation of China under Grant No.62301315, Startup Fund for Young Faculty at SJTU (SFYF at SJTU) under Grant No.23X010501967 and Shanghai Municipal Health Commission Health Industry Clinical Research Special Project under Grant No.202340010.
\item \textbf{Conflict of interest} \\  The authors declare that they have no conflict of interest.
\item \textbf{Ethics approval} \\  Not Applicable. The datasets and the work do not contain personal or sensitive information, no ethical issue is concerned.
\item \textbf{Consent to participate} \\  The authors are fine that the work is submitted and published by Machine Learning Journal. There is no human study in this work, so this aspect is not applicable.
\item \textbf{Consent for publication} \\  The authors are fine that the work (including all content, data and images) is published by Machine Learning Journal.
\item \textbf{Availability of data and material} \\  The data used for the experiments in this paper are available online, see Section~\ref{sec4.1} for more details.
\item \textbf{Code availability} \\  The code is publicly available at \href{https://github.com/JCruan519/MEW-UNet}{https://github.com/JCruan519/MEW-UNet}.
\item \textbf{Authors' contributions} \\  Suncheng Xiang and Jiacheng Ruan contributed conception and design of the study, as well as the experimental process and interpreted model results. Suncheng Xiang obtained funding for the project and provided clinical guidance. Suncheng Xiang drafted the manuscript. All authors contributed to manuscript revision, read and approved the submitted version.
\end{itemize}


\bibliography{sn-bibliography}

\begin{thebibliography}{49}
\providecommand{\natexlab}[1]{#1}
\providecommand{\url}[1]{{#1}}
\providecommand{\urlprefix}{URL }
\providecommand{\doi}[1]{\url{https://doi.org/#1}}
\providecommand{\eprint}[2][]{\url{#2}}
 \bibcommenthead

\bibitem[{Bernard et~al(2018)Bernard, Lalande, Zotti, Cervenansky, Yang, Heng, Cetin, Lekadir, Camara, Ballester et~al}]{acdc}
Bernard O, Lalande A, Zotti C, et~al (2018) Deep learning techniques for automatic mri cardiac multi-structures segmentation and diagnosis: is the problem solved? IEEE transactions on medical imaging 37(11):2514--2525

\bibitem[{Berseth(2017)}]{isic17}
Berseth M (2017) Isic 2017-skin lesion analysis towards melanoma detection. arXiv preprint arXiv:170300523

\bibitem[{Cao et~al(2021)Cao, Wang, Chen, Jiang, Zhang, Tian, and Wang}]{swinunet}
Cao H, Wang Y, Chen J, et~al (2021) Swin-unet: Unet-like pure transformer for medical image segmentation. arXiv preprint arXiv:210505537

\bibitem[{Chen et~al(2021)Chen, Lu, Yu, Luo, Adeli, Wang, Lu, Yuille, and Zhou}]{transunet}
Chen J, Lu Y, Yu Q, et~al (2021) Transunet: Transformers make strong encoders for medical image segmentation. arXiv preprint arXiv:210204306

\bibitem[{Chen et~al(2022)Chen, He, Zhuo, Ma, Ha, and Chan}]{tvconv}
Chen J, He T, Zhuo W, et~al (2022) Tvconv: Efficient translation variant convolution for layout-aware visual processing. In: Proceedings of the IEEE/CVF Conference on Computer Vision and Pattern Recognition, pp 12,548--12,558

\bibitem[{Chen et~al(2023)Chen, Cai, Cai, Yu, Qian, and Xiang}]{chen2023colo}
Chen Q, Cai S, Cai C, et~al (2023) Colo-scrl: Self-supervised contrastive representation learning for colonoscopic video retrieval. arXiv preprint arXiv:230315671

\bibitem[{Codella et~al(2019)Codella, Rotemberg, Tschandl, Celebi, Dusza, Gutman, Helba, Kalloo, Liopyris, Marchetti et~al}]{isic18}
Codella N, Rotemberg V, Tschandl P, et~al (2019) Skin lesion analysis toward melanoma detection 2018: A challenge hosted by the international skin imaging collaboration (isic). arXiv preprint arXiv:190203368

\bibitem[{Dolz et~al(2018)Dolz, Gopinath, Yuan, Lombaert, Desrosiers, and Ayed}]{dolz2018hyperdense}
Dolz J, Gopinath K, Yuan J, et~al (2018) Hyperdense-net: a hyper-densely connected cnn for multi-modal image segmentation. IEEE transactions on medical imaging 38(5):1116--1126

\bibitem[{Dosovitskiy et~al(2020{\natexlab{a}})Dosovitskiy, Beyer, Kolesnikov, Weissenborn, Zhai, Unterthiner, Dehghani, Minderer, Heigold, Gelly et~al}]{vit}
Dosovitskiy A, Beyer L, Kolesnikov A, et~al (2020{\natexlab{a}}) An image is worth 16x16 words: Transformers for image recognition at scale. arXiv preprint arXiv:201011929

\bibitem[{Dosovitskiy et~al(2020{\natexlab{b}})Dosovitskiy, Beyer, Kolesnikov, Weissenborn, Zhai, Unterthiner, Dehghani, Minderer, Heigold, Gelly et~al}]{dosovitskiy2020image}
Dosovitskiy A, Beyer L, Kolesnikov A, et~al (2020{\natexlab{b}}) An image is worth 16x16 words: Transformers for image recognition at scale. arXiv preprint arXiv:201011929

\bibitem[{Duan et~al(2023)Duan, Long, Wang, Zhang, Willcocks, and Shao}]{duan2023dynamic}
Duan H, Long Y, Wang S, et~al (2023) Dynamic unary convolution in transformers. IEEE Transactions on Pattern Analysis and Machine Intelligence

\bibitem[{Duan et~al(2024)Duan, Wan, Sun, Huang, Wang, Xin, Long, and Zheng}]{duan2024wearable}
Duan H, Wan F, Sun R, et~al (2024) Wearable-based behaviour interpolation for semi-supervised human activity recognition. Available at SSRN 4342069

\bibitem[{Gao et~al(2023)Gao, Ruan, Xiang, Yu, Ji, Xie, Liu, and Fu}]{gao2023lamm}
Gao J, Ruan J, Xiang S, et~al (2023) Lamm: Label alignment for multi-modal prompt learning. arXiv preprint arXiv:231208212

\bibitem[{Gao et~al(2024)Gao, Ruan, Xiang, Yu, Ji, Xie, Liu, and Fu}]{gao2024lamm}
Gao J, Ruan J, Xiang S, et~al (2024) Lamm: Label alignment for multi-modal prompt learning. In: Proceedings of the AAAI Conference on Artificial Intelligence, pp 1815--1823

\bibitem[{Gao et~al(2022)Gao, Zhou, Liu, and Metaxas}]{utnetv2}
Gao Y, Zhou M, Liu D, et~al (2022) A multi-scale transformer for medical image segmentation: Architectures, model efficiency, and benchmarks. arXiv preprint arXiv:220300131

\bibitem[{Huang et~al(2021)Huang, Zhou, Chen, Chen, and Lan}]{bibmMedicalFrequencyDomainLearning}
Huang Y, Zhou C, Chen L, et~al (2021) Medical frequency domain learning: Consider inter-class and intra-class frequency for medical image segmentation and classification. In: 2021 IEEE International Conference on Bioinformatics and Biomedicine (BIBM), IEEE, pp 897--904

\bibitem[{Ioffe and Szegedy(2015)}]{bn}
Ioffe S, Szegedy C (2015) Batch normalization: Accelerating deep network training by reducing internal covariate shift. In: International conference on machine learning, pmlr, pp 448--456

\bibitem[{Isensee et~al(2021)Isensee, Jaeger, Kohl, Petersen, and Maier-Hein}]{isensee2021nnu}
Isensee F, Jaeger PF, Kohl SA, et~al (2021) nnu-net: a self-configuring method for deep learning-based biomedical image segmentation. Nature methods 18(2):203--211

\bibitem[{Kamnitsas et~al(2017)Kamnitsas, Ledig, Newcombe, Simpson, Kane, Menon, Rueckert, and Glocker}]{kamnitsas2017efficient}
Kamnitsas K, Ledig C, Newcombe VF, et~al (2017) Efficient multi-scale 3d cnn with fully connected crf for accurate brain lesion segmentation. Medical image analysis 36:61--78

\bibitem[{Karimi et~al(2021)Karimi, Vasylechko, and Gholipour}]{karimi2021convolution}
Karimi D, Vasylechko SD, Gholipour A (2021) Convolution-free medical image segmentation using transformers. In: Medical Image Computing and Computer Assisted Intervention--MICCAI 2021: 24th International Conference, Strasbourg, France, September 27--October 1, 2021, Proceedings, Part I 24, Springer, pp 78--88

\bibitem[{Landman et~al(2015)Landman, Xu, Igelsias, Styner, Langerak, and Klein}]{synapse}
Landman B, Xu Z, Igelsias J, et~al (2015) Miccai multi-atlas labeling beyond the cranial vault--workshop and challenge. In: Proc. MICCAI Multi-Atlas Labeling Beyond Cranial Vault—Workshop Challenge, p~12

\bibitem[{Li et~al(2023)Li, Yu, Xiang, Liu, and Fu}]{li2023av}
Li Y, Yu Z, Xiang S, et~al (2023) Av-tad: Audio-visual temporal action detection with transformer. In: ICASSP 2023-2023 IEEE International Conference on Acoustics, Speech and Signal Processing (ICASSP), IEEE, pp 1--5

\bibitem[{Loshchilov and Hutter(2016)}]{cosineannealingLR}
Loshchilov I, Hutter F (2016) Sgdr: Stochastic gradient descent with warm restarts. arXiv preprint arXiv:160803983

\bibitem[{Loshchilov and Hutter(2017)}]{adamw}
Loshchilov I, Hutter F (2017) Decoupled weight decay regularization. arXiv preprint arXiv:171105101

\bibitem[{Milletari et~al(2016)Milletari, Navab, and Ahmadi}]{vnet}
Milletari F, Navab N, Ahmadi SA (2016) V-net: Fully convolutional neural networks for volumetric medical image segmentation. In: 2016 fourth international conference on 3D vision (3DV), IEEE, pp 565--571

\bibitem[{Oktay et~al(2018)Oktay, Schlemper, Folgoc, Lee, Heinrich, Misawa, Mori, McDonagh, Hammerla, Kainz et~al}]{attentionunet}
Oktay O, Schlemper J, Folgoc LL, et~al (2018) Attention u-net: Learning where to look for the pancreas. arXiv preprint arXiv:180403999

\bibitem[{Rao et~al(2021)Rao, Zhao, Zhu, Lu, and Zhou}]{gfnet}
Rao Y, Zhao W, Zhu Z, et~al (2021) Global filter networks for image classification. Advances in Neural Information Processing Systems 34:980--993

\bibitem[{Ronneberger et~al(2015)Ronneberger, Fischer, and Brox}]{unet}
Ronneberger O, Fischer P, Brox T (2015) U-net: Convolutional networks for biomedical image segmentation. In: International Conference on Medical image computing and computer-assisted intervention, Springer, pp 234--241

\bibitem[{Ruan and Xiang(2024)}]{ruan2024vm}
Ruan J, Xiang S (2024) Vm-unet: Vision mamba unet for medical image segmentation. arXiv preprint arXiv:240202491

\bibitem[{Ruan et~al(2022)Ruan, Xiang, Xie, Liu, and Fu}]{malunet}
Ruan J, Xiang S, Xie M, et~al (2022) Malunet: A multi-attention and light-weight unet for skin lesion segmentation. In: 2022 IEEE International Conference on Bioinformatics and Biomedicine (BIBM), IEEE, pp 1150--1156

\bibitem[{Ruan et~al(2023)Ruan, Xie, Gao, Liu, and Fu}]{egeunet}
Ruan J, Xie M, Gao J, et~al (2023) Ege-unet: an efficient group enhanced unet for skin lesion segmentation. In: International Conference on Medical Image Computing and Computer-Assisted Intervention, Springer, pp 481--490

\bibitem[{Sandler et~al(2018)Sandler, Howard, Zhu, Zhmoginov, and Chen}]{mobilenetv2}
Sandler M, Howard A, Zhu M, et~al (2018) Mobilenetv2: Inverted residuals and linear bottlenecks. In: Proceedings of the IEEE conference on computer vision and pattern recognition, pp 4510--4520

\bibitem[{Wang et~al(2022{\natexlab{a}})Wang, Cao, Wang, and Zaiane}]{uctransnet}
Wang H, Cao P, Wang J, et~al (2022{\natexlab{a}}) Uctransnet: rethinking the skip connections in u-net from a channel-wise perspective with transformer. In: Proceedings of the AAAI Conference on Artificial Intelligence, pp 2441--2449

\bibitem[{Wang et~al(2022{\natexlab{b}})Wang, Cao, Wang, and Zaiane}]{wang2022uctransnet}
Wang H, Cao P, Wang J, et~al (2022{\natexlab{b}}) Uctransnet: rethinking the skip connections in u-net from a channel-wise perspective with transformer. In: Proceedings of the AAAI conference on artificial intelligence, pp 2441--2449

\bibitem[{Wang et~al(2022{\natexlab{c}})Wang, Xie, Lin, Iwamoto, Han, Chen, and Tong}]{mtunet}
Wang H, Xie S, Lin L, et~al (2022{\natexlab{c}}) Mixed transformer u-net for medical image segmentation. In: ICASSP 2022-2022 IEEE International Conference on Acoustics, Speech and Signal Processing (ICASSP), IEEE, pp 2390--2394

\bibitem[{Wei et~al(2021)Wei, Hu, Zhang, Li, Zhou, and Cui}]{sanet}
Wei J, Hu Y, Zhang R, et~al (2021) Shallow attention network for polyp segmentation. In: International Conference on Medical Image Computing and Computer-Assisted Intervention, Springer, pp 699--708

\bibitem[{Wu and He(2018)}]{gn}
Wu Y, He K (2018) Group normalization. In: Proceedings of the European conference on computer vision (ECCV), pp 3--19

\bibitem[{Xiang et~al(2020{\natexlab{a}})Xiang, Fu, and Liu}]{xiang2020progressive}
Xiang S, Fu Y, Liu T (2020{\natexlab{a}}) Progressive learning with style transfer for distant domain adaptation. IET Image Processing 14(14):3527--3535

\bibitem[{Xiang et~al(2020{\natexlab{b}})Xiang, Fu, You, and Liu}]{xiang2020unsupervised1}
Xiang S, Fu Y, You G, et~al (2020{\natexlab{b}}) Unsupervised domain adaptation through synthesis for person re-identification. In: 2020 IEEE International Conference on Multimedia and Expo (ICME), IEEE, pp 1--6

\bibitem[{Xiang et~al(2022)Xiang, You, Li, Guan, Liu, Qian, and Fu}]{xiang2022rethinking}
Xiang S, You G, Li L, et~al (2022) Rethinking illumination for person re-identification: A unified view. In: Proceedings of the IEEE/CVF Conference on Computer Vision and Pattern Recognition, pp 4731--4739

\bibitem[{Xiang et~al(2023{\natexlab{a}})Xiang, Chen, Ran, Yu, Liu, Qian, and Fu}]{xiang2023deep}
Xiang S, Chen H, Ran W, et~al (2023{\natexlab{a}}) Deep multimodal representation learning for generalizable person re-identification. Machine Learning pp 1--19

\bibitem[{Xiang et~al(2023{\natexlab{b}})Xiang, Fu, Guan, and Liu}]{xiang2023learning}
Xiang S, Fu Y, Guan M, et~al (2023{\natexlab{b}}) Learning from self-discrepancy via multiple co-teaching for cross-domain person re-identification. Machine Learning 112(6):1923--1940

\bibitem[{Xiang et~al(2023{\natexlab{c}})Xiang, Qian, Gao, Zhang, Liu, and Fu}]{xiang2023rethinking}
Xiang S, Qian D, Gao J, et~al (2023{\natexlab{c}}) Rethinking person re-identification via semantic-based pretraining. ACM Transactions on Multimedia Computing, Communications and Applications 20(3):1--17

\bibitem[{Xiang et~al(2023{\natexlab{d}})Xiang, Qian, Guan, Yan, Liu, Fu, and You}]{xiang2023less}
Xiang S, Qian D, Guan M, et~al (2023{\natexlab{d}}) Less is more: Learning from synthetic data with fine-grained attributes for person re-identification. ACM Transactions on Multimedia Computing, Communications and Applications 19(5s):1--20

\bibitem[{You et~al(2023)You, Xia, Chen, Wu, Xiang, and Wang}]{you2023autokary2022}
You D, Xia P, Chen Q, et~al (2023) Autokary2022: A large-scale densely annotated dateset for chromosome instance segmentation. arXiv preprint arXiv:230315839

\bibitem[{Yushkevich et~al(2006)Yushkevich, Piven, Cody~Hazlett, Gimpel~Smith, Ho, Gee, and Gerig}]{ITK-SNAP}
Yushkevich PA, Piven J, Cody~Hazlett H, et~al (2006) User-guided {3D} active contour segmentation of anatomical structures: Significantly improved efficiency and reliability. Neuroimage 31(3):1116--1128

\bibitem[{Zhang et~al(2021)Zhang, Liu, and Hu}]{transfuse}
Zhang Y, Liu H, Hu Q (2021) Transfuse: Fusing transformers and cnns for medical image segmentation. In: International Conference on Medical Image Computing and Computer-Assisted Intervention, Springer, pp 14--24

\bibitem[{Zhong et~al(2022)Zhong, Li, Tang, Kuang, Wu, and Ding}]{frequencyCOD}
Zhong Y, Li B, Tang L, et~al (2022) Detecting camouflaged object in frequency domain. In: Proceedings of the IEEE/CVF Conference on Computer Vision and Pattern Recognition, pp 4504--4513

\bibitem[{Zhou et~al(2018)Zhou, Rahman~Siddiquee, Tajbakhsh, and Liang}]{unet++}
Zhou Z, Rahman~Siddiquee MM, Tajbakhsh N, et~al (2018) Unet++: A nested u-net architecture for medical image segmentation. In: Deep learning in medical image analysis and multimodal learning for clinical decision support. Springer, p 3--11

\end{thebibliography}


\end{document}